\pdfoutput=1
\documentclass[12pt,a4paper]{article}

\usepackage{ifthen}

\newboolean{pdflatex}
\setboolean{pdflatex}{true} 

\newboolean{articletitles}
\setboolean{articletitles}{true} 

\usepackage{microtype}
\usepackage{lineno}  
\usepackage{xspace} 

\usepackage{graphicx}  
\usepackage{color}
\usepackage{colortbl}

\usepackage{amsmath} 
\usepackage{amssymb}
\usepackage{amsfonts}
\usepackage{upgreek} 

\newcommand*\patchAmsMathEnvironmentForLineno[1]{%
\expandafter\let\csname old#1\expandafter\endcsname\csname #1\endcsname
\expandafter\let\csname oldend#1\expandafter\endcsname\csname
end#1\endcsname
 \renewenvironment{#1}%
   {\linenomath\csname old#1\endcsname}%
   {\csname oldend#1\endcsname\endlinenomath}%
}
\newcommand*\patchBothAmsMathEnvironmentsForLineno[1]{%
  \patchAmsMathEnvironmentForLineno{#1}%
  \patchAmsMathEnvironmentForLineno{#1*}%
}
\AtBeginDocument{%
\patchBothAmsMathEnvironmentsForLineno{equation}%
\patchBothAmsMathEnvironmentsForLineno{align}%
\patchBothAmsMathEnvironmentsForLineno{flalign}%
\patchBothAmsMathEnvironmentsForLineno{alignat}%
\patchBothAmsMathEnvironmentsForLineno{gather}%
\patchBothAmsMathEnvironmentsForLineno{multline}%
}

\usepackage{hyperref}    
\usepackage[all]{hypcap} 

\def\lhcb {\mbox{LHCb}\xspace}
\def\lhc    {\mbox{LHC}\xspace}

\def\velo   {VELO\xspace}

\def\Pmu         {\ensuremath{\mu}\xspace}                 
\def\Ppi         {\ensuremath{\pi}\xspace}                 
\def\Ppsi        {\ensuremath{\psi}\xspace}                 
\mathchardef\PLambda="7103
\def\PB      {\ensuremath{B}\xspace}                 
\def\PJ      {\ensuremath{J}\xspace}                 
\def\PK      {\ensuremath{K}\xspace}                 
\def\PZ      {\ensuremath{Z}\xspace}                 
\def\Pb      {\ensuremath{b}\xspace}                 
\def\Pc      {\ensuremath{c}\xspace}                 
\def\Pe      {\ensuremath{e}\xspace}                 
\def\Pp      {\ensuremath{p}\xspace}

\def\en         {\ensuremath{\Pe^-}\xspace}   
\def\ep         {\ensuremath{\Pe^+}\xspace}

\def\mup        {\ensuremath{\Pmu^+}\xspace}
\def\mun        {\ensuremath{\Pmu^-}\xspace} 

\def\Z      {\ensuremath{\PZ^0}\xspace}

\def\cquark    {\ensuremath{\Pc}\xspace}

\def\bquark    {\ensuremath{\Pb}\xspace}
\def\bquarkbar {\ensuremath{\overline \bquark}\xspace}

\def\pion  {\ensuremath{\Ppi}\xspace}

\def\pim   {\ensuremath{\pion^-}\xspace}

\def\kaon  {\ensuremath{\PK}\xspace}
\def\KS    {\ensuremath{\kaon^0_{\rm\scriptscriptstyle S}}\xspace} 

\def\B       {\ensuremath{\PB}\xspace}
\def\Bz      {\ensuremath{\B^0}\xspace}
\def\Bd      {\ensuremath{\B^0}\xspace}

\def\jpsi     {\ensuremath{{\PJ\mskip -3mu/\mskip -2mu\Ppsi\mskip 2mu}}\xspace}

\def\proton      {\ensuremath{\Pp}\xspace}

\def\L {\ensuremath{\PLambda}\xspace}
\def\Lbar {\ensuremath{\kern 0.1em\overline{\kern -0.1em\PLambda}}\xspace}

\def\Lb      {\ensuremath{\L^0_\bquark}\xspace}
\def\Lbbar   {\ensuremath{\Lbar^0_\bquark}\xspace}

\newcommand{\decay}[2]{\ensuremath{#1\!\to #2}\xspace}         

\def\CP                {\ensuremath{C\!P}\xspace}

\newcommand{\tev}{\ensuremath{\mathrm{\,Te\kern -0.1em V}}\xspace}
\newcommand{\gev}{\ensuremath{\mathrm{\,Ge\kern -0.1em V}}\xspace}
\newcommand{\mev}{\ensuremath{\mathrm{\,Me\kern -0.1em V}}\xspace}
\newcommand{\kev}{\ensuremath{\mathrm{\,ke\kern -0.1em V}}\xspace}
\newcommand{\ev}{\ensuremath{\mathrm{\,e\kern -0.1em V}}\xspace}
\newcommand{\gevc}{\ensuremath{{\mathrm{\,Ge\kern -0.1em V\!/}c}}\xspace}
\newcommand{\mevc}{\ensuremath{{\mathrm{\,Me\kern -0.1em V\!/}c}}\xspace}
\newcommand{\gevcc}{\ensuremath{{\mathrm{\,Ge\kern -0.1em V\!/}c^2}}\xspace}
\newcommand{\gevgevcccc}{\ensuremath{{\mathrm{\,Ge\kern -0.1em V^2\!/}c^4}}\xspace}
\newcommand{\mevcc}{\ensuremath{{\mathrm{\,Me\kern -0.1em V\!/}c^2}}\xspace}

\def\mum  {\ensuremath{\,\upmu\rm m}\xspace}
\def\invfb   {\ensuremath{\mbox{\,fb}^{-1}}\xspace}

\newcommand{\chisq}{\ensuremath{\chi^2}\xspace}

\def\deriv {\ensuremath{\mathrm{d}}}

\def\PDF {PDF\xspace}

\def\sPlot{\mbox{\em sPlot}\xspace}
\def\sWeight{\mbox{\em sWeight}\xspace}

\def\sqs   {\ensuremath{\protect\sqrt{s}}\xspace}

\def\pt         {\mbox{$p_{\rm T}$}\xspace}

\def\evtgen     {\mbox{\textsc{EvtGen}}\xspace}
\def\pythia     {\mbox{\textsc{Pythia}}\xspace}
\def\geant      {\mbox{\textsc{Geant4}}\xspace}

\def\photos     {\mbox{\textsc{Photos}}\xspace}

\newcommand{\ie}{\mbox{\itshape i.e.}}

\def\LbToLJPsi    {\decay{\Lb}{\jpsi\L}}
\def\LbToLppiJPsimumu    {\decay{\Lb}{\jpsi(\to \mup \mun)\L(\to \proton \pim)}}
\def\LToppi{\decay{\L}{\proton \pim}}

\def\Lst{\ensuremath{\L^{*}}\xspace}
\def\LbToLgamma    {\decay{\Lb}{\L\gamma}}
\def\LbToLstgamma    {\decay{\Lb}{\Lst\gamma}}

\def\BdToJPsiKS  {\decay{\Bd}{\jpsi\KS}}

\def\Polb{\ensuremath{P_b}\xspace}
\def\ab{\ensuremath{\alpha_b}\xspace}
\def\al{\ensuremath{\alpha_{\L}}\xspace}
\def\albar{\ensuremath{\alpha_{\Lbar}}\xspace}
\def\rz{\ensuremath{r_0}\xspace}
\def\ro{\ensuremath{r_1}\xspace}
\def\thz{\ensuremath{\theta}\xspace}
\def\tho{\ensuremath{\theta_{1}}\xspace}
\def\tht{\ensuremath{\theta_{2}}\xspace}

\def\phio{\ensuremath{\phi_{1}}\xspace}
\def\phit{\ensuremath{\phi_{2}}\xspace}

\def\cthz{\ensuremath{\cos\thz}\xspace}
\def\ctho{\ensuremath{\cos\tho}\xspace}
\def\ctht{\ensuremath{\cos\tht}\xspace}
\def\cttht{\ensuremath{\cos^2\tht}\xspace}

\def\half{\ensuremath{\frac{1}{2}}\xspace}

\def\mll{\ensuremath{\mathcal{M}_{\lambda_1\lambda_2}}\xspace}

\def\map{\ensuremath{\mathcal{M}_{+\half,0}}\xspace}
\def\mam{\ensuremath{\mathcal{M}_{-\half,0}}\xspace}
\def\mbp{\ensuremath{\mathcal{M}_{-\half,-1}}\xspace}
\def\mbm{\ensuremath{\mathcal{M}_{+\half,+1}}\xspace}

\def\mapt{\ensuremath{|\map|^2}\xspace}
\def\mamt{\ensuremath{|\mam|^2}\xspace}
\def\mbpt{\ensuremath{|\mbp|^2}\xspace}
\def\mbmt{\ensuremath{|\mbm|^2}\xspace}

\def\wmass{\ensuremath{w_{\rm mass}}\xspace}
\def\wacc{\ensuremath{w_{\rm acc}}\xspace}
\def\wtot{\ensuremath{w_{\rm tot}}\xspace}

\def\xf{\ensuremath{x_{\rm F}}\xspace}
\def\pl{\mbox{$p_{\rm L}$}\xspace}

\def\PDFs {PDFs\xspace}

\def\dGdomegat{\ensuremath{\frac{\deriv\Gamma}{\deriv\Omega_3}}}
\def\dGdomegaf{\ensuremath{\frac{\deriv\Gamma}{\deriv\Omega_5}}}

\def\ipchisq{\ensuremath{\chisq_{\rm IP}}\xspace}

\usepackage{cite} 
\usepackage{mciteplus}

\usepackage{mathtools} 

\begin{document}

\renewcommand{\thefootnote}{\fnsymbol{footnote}}
\setcounter{footnote}{1}

\begin{titlepage}
\pagenumbering{roman}

\vspace*{-1.5cm}
\centerline{\large EUROPEAN ORGANIZATION FOR NUCLEAR RESEARCH (CERN)}
\vspace*{1.5cm}
\hspace*{-0.5cm}
\begin{tabular*}{\linewidth}{lc@{\extracolsep{\fill}}r}
\ifthenelse{\boolean{pdflatex}}
{\vspace*{-2.7cm}\mbox{\!\!\!\includegraphics[width=.14\textwidth]{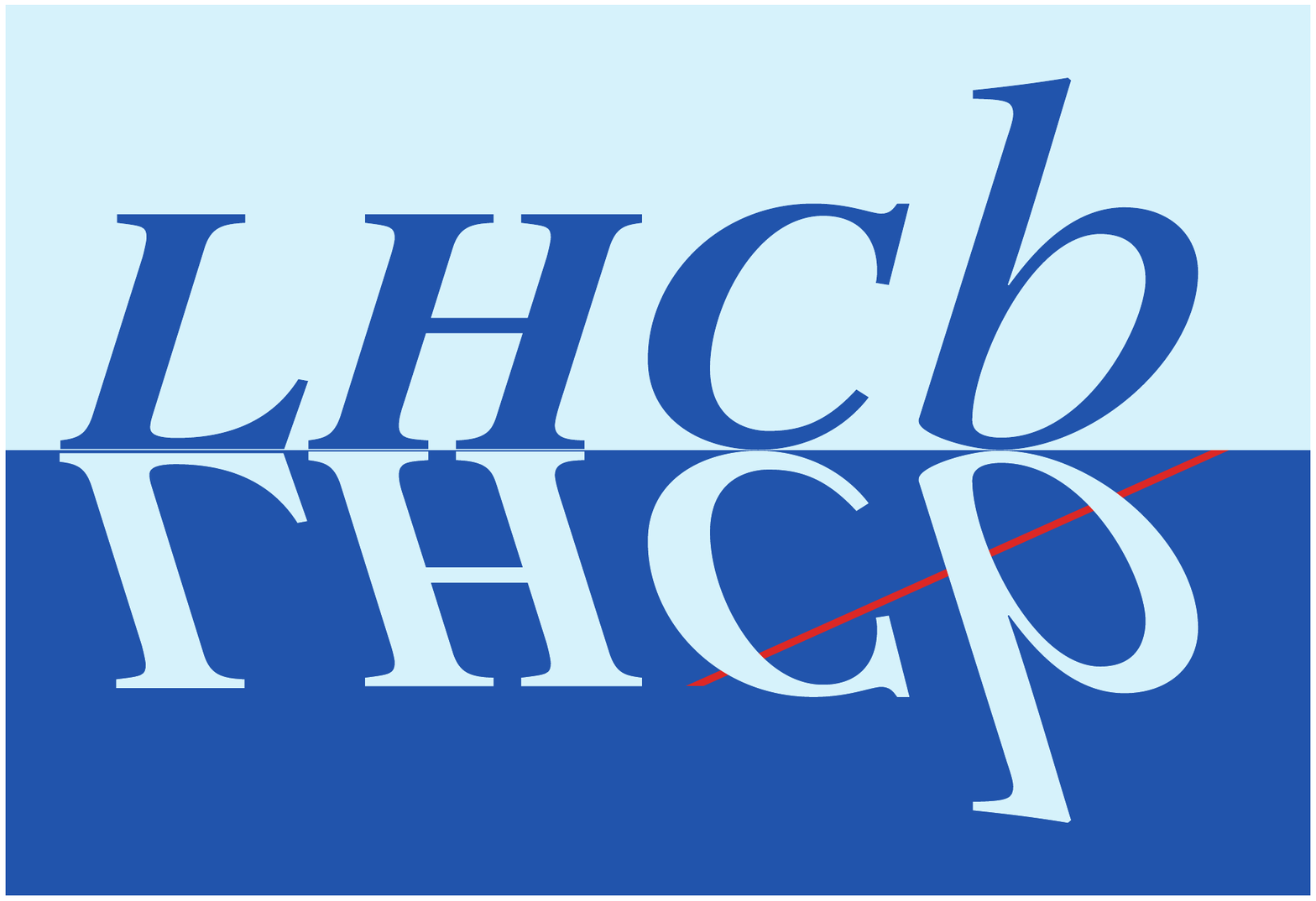}} & &}%
{\vspace*{-1.2cm}\mbox{\!\!\!\includegraphics[width=.12\textwidth]{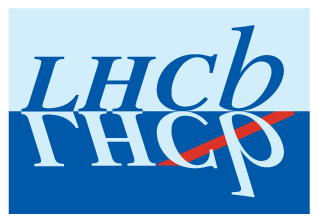}} & &}%
\\
 & & CERN-PH-EP-2013-020 \\  
 & & LHCb-PAPER-2012-057  \\  
 & & June 13, 2013 \\ 
\end{tabular*}

\vspace*{3.0cm}

{\bf\boldmath\huge
\begin{center}
Measurements of the \LbToLJPsi decay amplitudes and the \Lb polarisation in $\proton\proton$ collisions at $\sqs = 7 \tev$
\end{center}
}

\vspace*{2.0cm}

\begin{center}
The LHCb collaboration\footnote{Authors are listed on the following pages.}
\end{center}

\vspace{\fill}

\begin{abstract}
  \noindent
  An angular analysis of \LbToLJPsi decays is performed using a data sample corresponding to $1.0 \invfb$ collected in $\proton\proton$ collisions at $\sqs = 7 \tev$ with the \lhcb detector at the LHC. A parity violating asymmetry parameter characterising the \LbToLJPsi decay of $0.05 \pm 0.17 \pm 0.07$ and a \Lb transverse production polarisation of $0.06 \pm 0.07 \pm 0.02$ are measured, where the first uncertainty is statistical and the second systematic.
\end{abstract}

\vspace*{2.0cm}

\begin{center}
Submitted to Physics Letters B
\end{center}

\vspace{\fill}

{\footnotesize 
\centerline{\copyright~CERN on behalf of the \lhcb collaboration, license \href{http://creativecommons.org/licenses/by/3.0/}{CC-BY-3.0}.}}
\vspace*{2mm}

\addcontentsline{toc}{section}{Title}

\end{titlepage}


\newpage
\setcounter{page}{2}
\mbox{~}
\newpage

\centerline{\large\bf LHCb collaboration}
\begin{flushleft}
\small
R.~Aaij$^{40}$, 
C.~Abellan~Beteta$^{35,n}$, 
B.~Adeva$^{36}$, 
M.~Adinolfi$^{45}$, 
C.~Adrover$^{6}$, 
A.~Affolder$^{51}$, 
Z.~Ajaltouni$^{5}$, 
J.~Albrecht$^{9}$, 
F.~Alessio$^{37}$, 
M.~Alexander$^{50}$
% ...

{\footnotesize \it
% ...
$ ^{5}$Clermont Universit\'{e}, Universit\'{e} Blaise Pascal, CNRS/IN2P3, LPC, Clermont-Ferrand, France\\
$ ^{6}$CPPM, Aix-Marseille Universit\'{e}, CNRS/IN2P3, Marseille, France\\
% ...
$ ^{9}$Fakult\"{a}t Physik, Technische Universit\"{a}t Dortmund, Dortmund, Germany\\
% ...
$ ^{35}$Universitat de Barcelona, Barcelona, Spain\\
$ ^{36}$Universidad de Santiago de Compostela, Santiago de Compostela, Spain\\
$ ^{37}$European Organization for Nuclear Research (CERN), Geneva, Switzerland\\
% ...
$ ^{40}$Nikhef National Institute for Subatomic Physics, Amsterdam, The Netherlands\\
% ...
$ ^{45}$H.H. Wills Physics Laboratory, University of Bristol, Bristol, United Kingdom\\
% ...
$ ^{50}$School of Physics and Astronomy, University of Glasgow, Glasgow, United Kingdom\\
$ ^{51}$Oliver Lodge Laboratory, University of Liverpool, Liverpool, United Kingdom\\
% ...
$ ^{n}$LIFAELS, La Salle, Universitat Ramon Llull, Barcelona, Spain\\
% ...
}
\end{flushleft}

\cleardoublepage

\renewcommand{\thefootnote}{\arabic{footnote}}
\setcounter{footnote}{0}


\pagestyle{plain} 
\setcounter{page}{1}
\pagenumbering{arabic}


\section{Introduction}
\label{sec:Introduction}

For \Lb baryons originating from energetic \bquark-quarks, heavy-quark effective theory (HQET) predicts a large fraction of the transverse \bquark-quark polarisation to be retained after hadronisation~\cite{Mannel:1991bs,Falk:1993rf}, while the longitudinal polarisation should vanish due to parity conservation in strong interactions. For \Lb baryons produced in $\en\ep \to \Z \to \bquark \bquarkbar$ transitions, a substantial polarisation is measured~\cite{Buskulic:1995mf,Abbiendi:1998uz,Abreu:1999gf}, in agreement with the $\Z\bquark\bquarkbar$ coupling of the Standard Model~(SM). There is no previous polarisation measurement for \Lb baryons produced at hadron colliders. The transverse polarisation is estimated to be $\mathcal{O}(10\%)$ in Ref.~\cite{Hiller:2007ur} while Ref.~\cite{Ajaltouni:2004zu} mentions it could be as large as 20\%. However, for \L baryons produced in fixed-target experiments~\cite{Ramberg:1994tk,Fanti:1998px,Abt:2006da}, the polarisation was observed to depend strongly on the Feynman variable $\xf = 2 \, \pl/\sqs$, \pl being the \L longitudinal momentum and $\sqs$ the collision centre-of-mass energy, and to vanish at $\xf \approx 0$. Extrapolating these results and taking into account the very small $\xf \approx 0.02$ value for \Lb produced at the Large Hadron Collider (\lhc) at $\sqs = 7 \tev$, this could imply a polarisation much smaller than $10\%$.

In this Letter, we perform an angular analysis of \LbToLppiJPsimumu decays using $1.0 \invfb$ of $\proton\proton$ collision data collected in 2011 with the \lhcb detector~\cite{Alves:2008zz} at the \lhc at $\sqs = 7 \tev$. Owing to the well-measured \LToppi decay asymmetry parameter (\al)~\cite{PDG2012} and the known behaviour of the decay of a vector particle into two leptons, the final state angular distribution contains sufficient information to measure the \Lb production polarisation and the decay amplitudes~\cite{Lednicky:1985zx}. The asymmetry of the \Lbar decay (\albar) is much less precisely measured~\cite{PDG2012}, however by neglecting possible \CP violation effects, which are predicted to be very small in the SM~\cite{Donoghue:1986hh,Donoghue:1986nn}, \al and $-\albar$ can be assumed to be equal. Similarly, \CP violation effects in \Lb decays are neglected, and the decay amplitudes of the \Lb and \Lbbar are therefore assumed to be equal. Inclusion of charge-conjugated modes is henceforth implied. The asymmetry parameter \ab in \LbToLJPsi decays, defined in Sec.~\ref{sec:angularform}, is calculated in many publications as summarised in Table~\ref{tab:alpha}. Most predictions lie in the range from $-21\%$ to $-10\%$ while Ref.~\cite{Ajaltouni:2004zu} obtains a large positive value using HQET. Note that the theoretical predictions depend on the calculations of the form-factors and experimental input that were available at the time they were made.

\begin{table}[b!]
  \caption{Theoretical predictions for the \LbToLJPsi decay asymmetry parameter \ab.}
  \begin{center}
    \begin{tabular}{lcc}
      Method & Value & Reference \\ \hline
      Factorisation & $-0.1$  &\cite{Cheng:1996cs} \\
      Factorisation & $-0.18$  &  \cite{Fayyazuddin:1998ap} \\
      Covariant oscillator quark model & $-0.208$ &  \cite{Mohanta:1998iu} \\
      Perturbative QCD & $-0.17$ to $-0.14$ & \cite{Chou:2001bn} \\
      Factorisation (HQET) & $0.777$   & \cite{Ajaltouni:2004zu} \\
      Light front quark model & $-0.204$ & \cite{Wei:2009np} \\
    \end{tabular}
  \end{center}
  \label{tab:alpha}
\end{table}

It should be noted that \Lb baryons can also be produced in the decay of heavier \bquark~baryons~\cite{Aaltonen:2007ar,CDF:2011ac,LHCb-PAPER-2012-012}, where the polarisation is partially diluted~\cite{Hiller:2007ur}. These strong decays are experimentally difficult to distinguish from \Lb that hadronise directly from a $\proton \proton$ collision and therefore contribute to the measurement presented in this study.

A sufficiently large \Lb polarisation would allow the photon helicity in \LbToLgamma and \LbToLstgamma decays to be probed~\cite{Mannel:1997xy,Legger:2006cq,Hiller:2007ur}. The photon helicity is sensitive to contributions from beyond the SM.

\section{Angular formalism}
\label{sec:angularform}

\begin{figure}[b!]
  \begin{center}
    \includegraphics[width=0.85\linewidth]{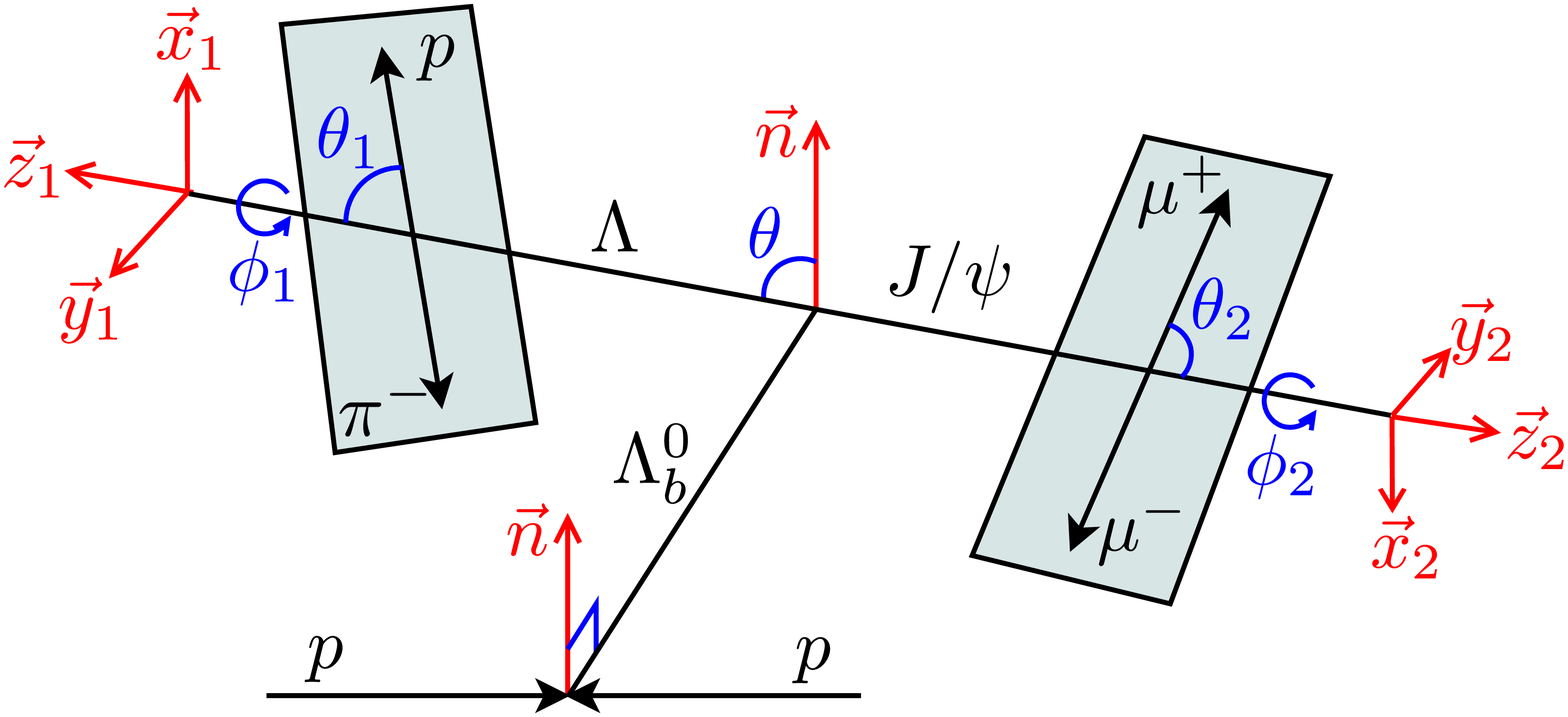}
    \vspace*{-0.5cm}
  \end{center}
  \caption{
    \small
    Definition of the five angles used to describe the \LbToLppiJPsimumu decay.}
  \label{fig:angles}
\end{figure}

The \Lb spin has not yet been measured but the quark model prediction is spin \half. The \LbToLJPsi mode is therefore the decay of a spin \half particle into a spin $1$ and a spin \half particle. In the helicity formalism, the decay can be described by four \mll helicity amplitudes (\map, \mam, \mbp\ and \mbm) where $\lambda_1$ ($\lambda_2$) is the helicity of the \L (\jpsi) particle. The angular distribution of the decay (${\dGdomegaf}$) is calculated in Ref.~\cite{Lednicky:1985zx} and reported in Ref.~\cite{Hrivnac:1994jx}. It depends on the five angles shown in Fig.~\ref{fig:angles}. The first angle, \thz, is the polar angle of the \L momentum in the \Lb rest-frame with respect to $\vec{n} = (\vec{p}_{\Lb} \times \vec{p}_{{\rm beam }})/ |\vec{p}_{\Lb} \times \vec{p}_{{\rm beam }}|$, a unit vector perpendicular to the production plane. The second and third angles are \tho and \phio, the polar and azimuthal angles of the proton in the \L rest-frame and calculated in the coordinate system defined by $\vec{z}_1 = \vec{p}_{\L}/|\vec{p}_{\L}|$ and $\vec{y}_1  = (\vec{n} \times \vec{p}_{\L}) / | \vec{n} \times \vec{p}_{\L} |$. The remaining angles are \tht and \phit, the polar and azimuthal angles of the positively-charged muon in the \jpsi rest-frame and calculated in the coordinate system defined by $\vec{z}_2 = \vec{p}_{\jpsi} / |\vec{p}_{\jpsi}|$ and $\vec{y}_2 = (\vec{n} \times \vec{p}_{\jpsi}) / | \vec{n} \times \vec{p}_{\jpsi}|$. The angular distribution also depends on the four \mll amplitudes, on the \al parameter, and on the transverse polarisation parameter \Polb, the projection of the \Lb polarisation vector on $\vec{n}$.

Assuming that the detector acceptance over \phio and \phit is uniformly distributed, the analysis can be simplified by integrating over the two azimuthal angles
\begin{align}
\label{eqn:w3}
  \dGdomegat(\cthz,\ctho,\ctht) & =  \int_{-\pi}^{\pi} \int_{-\pi}^{\pi} \; \dGdomegaf(\thz,\tho,\tht,\phio,\phit) \; \deriv\phio \; \deriv\phit \nonumber \\
                               & =  \frac{1}{16\pi}\sum_{i=0}^7 \; f_i(\mapt,\mamt,\mbpt,\mbmt) \nonumber \\
                               &  \hspace{2.2cm} g_i(\Polb,\al)  \; h_i(\cthz,\ctho,\ctht).
\end{align}

The functions describing the decay only depend on the magnitudes of the \mll amplitudes, on \Polb and \al, and on \cthz, \ctho, and \ctht. Using the normalisation condition ${\mapt + \mamt + \mbpt + \mbmt = 1}$, the $f_i$ functions can be written in terms of the following three parameters: ${\ab \equiv \mapt - \mamt + \mbpt -\mbmt}$, ${\rz \equiv \mapt + \mamt}$ and ${\ro \equiv \mapt - \mamt}$. The functions used to describe the angular distributions are shown in Table~\ref{tab:angdistrb3dim}. Four parameters (\Polb, \ab, \rz and \ro) have to be measured simultaneously from the angular distribution. The \ab parameter is the parity violating asymmetry characterising the \LbToLJPsi\ decay.

\begin{table}[tb!]
  \caption{Functions used to describe the angular distributions in three dimensions.}
  \begin{center}
    \begin{tabular}{llll}
      $i$ & $f_i(\ab,\rz,\ro)$ & $g_i(\Polb,\al)$ & $h_i(\cthz,\ctho,\ctht)$ \\ \hline \noalign{\vskip 1mm}
      0   & $1$                & $1$              & $1$    \\ \noalign{\vskip 1mm}
      1   & $\ab$              & $\Polb$          & $\cthz$  \\ \noalign{\vskip 1mm}
      2   & $2 \ro-\ab$         & $\al$            & $\ctho$  \\ \noalign{\vskip 1mm}
      3   & $2 \rz-1$           & $\Polb\al$       & $\cthz\ctho$  \\ \noalign{\vskip 1mm}
      4   & $\half(1-3 \rz)$    & $1$              & $\half(3\cttht-1)$     \\ \noalign{\vskip 1mm}
      5   & $\half(\ab-3 \ro)$  & $\Polb$          & $\half(3\cttht-1)\cthz$     \\ \noalign{\vskip 1mm}
      6   & $-\half (\ab + \ro)$  & $\al$            & $\half(3\cttht-1)\ctho$      \\ \noalign{\vskip 1mm}
      7   & $-\half (1 + \rz)$    & $\Polb\al$       & $\half(3\cttht-1)\cthz\ctho$      \\
    \end{tabular}
  \end{center}
  \label{tab:angdistrb3dim}
\end{table}

\section{Detector, trigger and simulation}
\label{sec:Detector}

The \lhcb detector~\cite{Alves:2008zz} is a single-arm forward
spectrometer covering the \mbox{pseudorapidity} range $2<\eta <5$,
designed for the study of particles containing \bquark or \cquark
quarks. The detector includes a high precision tracking system
consisting of a silicon-strip vertex detector (\velo) surrounding the $pp$
interaction region, a large-area silicon-strip detector located
upstream of a dipole magnet with a bending power of about
$4{\rm\,Tm}$, and three stations of silicon-strip detectors and straw
drift tubes placed downstream. The combined tracking system provides a momentum measurement with
relative uncertainty that varies from 0.4\% at 5\gevc to 0.6\% at 100\gevc,
and three-dimensional impact parameter (IP) resolution of 20\mum for
tracks with high transverse momentum. Charged hadrons are identified
using two ring-imaging Cherenkov detectors (RICH)~\cite{LHCb-DP-2012-003}. Photon, electron and
hadron candidates are identified by a calorimeter system consisting of
scintillating-pad and preshower detectors, an electromagnetic
calorimeter and a hadronic calorimeter. Muons are identified by a
system composed of alternating layers of iron and multiwire
proportional chambers~\cite{LHCb-DP-2012-002}.
The trigger~\cite{LHCb-DP-2012-004} consists of a
hardware stage, based on information from the calorimeter and muon
systems, followed by a software stage, which applies a full event
reconstruction.

The hardware trigger selects events containing a muon with a transverse momentum, $\pt>1.48\gevc$ or two muons with a product of their \pt larger than $(1.3\gevc)^2$. In the subsequent software trigger, we require two oppositely-charged muons having an invariant mass larger than $2800\mevcc$ and originating from the same vertex, or a single muon with $\pt>1.3\gevc$ and being significantly displaced with respect to all the primary $\proton\proton$ interaction vertices~(PVs) in the event, or a single muon with $p>10\gevc$ and $\pt>1.7\gevc$. Displaced muons are identified by means of their IP and \ipchisq, where the \ipchisq is the \chisq difference when the PV is fitted with or without the muon track. Finally, we require two oppositely-charged muons with an invariant mass within $120\mevcc$ of the nominal \jpsi mass~\cite{PDG2012} forming a common vertex which is significantly displaced from the PVs. Displaced \jpsi vertices are identified by computing the vertex separation \chisq, the \chisq difference between the PV and the \jpsi vertex. In the \LbToLJPsi selection described below, we use the muon pairs selected by the trigger.

Simulation is used to understand the detector efficiencies and resolutions and to train the analysis procedure. Proton-proton collisions are generated using \pythia~6.4~\cite{Sjostrand:2006za} with a specific \lhcb configuration~\cite{LHCb-PROC-2010-056}. Decays of hadronic particles are described by \evtgen~\cite{Lange:2001uf} in which final state radiation is generated using \photos~\cite{Golonka:2005pn}. The interaction of the generated particles with the detector and its response are implemented using the \geant toolkit~\cite{Allison:2006ve,*Agostinelli:2002hh} as described in Ref.~\cite{LHCb-PROC-2011-006}.

\section{Signal selection and background rejection}
\label{sec:selection}

A first set of loose requirements is applied to select \LbToLJPsi decays. Charged tracks are identified as either protons or pions using information provided by the RICH system. Candidate \L baryons are reconstructed from oppositely-charged proton and pion candidates. They are reconstructed either when the \L decays within the \velo (``long \L''), or when the decay occurs outside the \velo acceptance (``downstream \L''). The latter category increases the acceptance significantly for long-lived \L decays. In both cases, the two tracks are required to have $p > 2 \gevc$, to be well separated from the PVs and to originate from a common vertex. In addition, protons are required to have $\pt > 0.5 \gevc$ and pions to have $\pt > 0.1 \gevc$. Finally, the invariant mass of the \L candidates is required to be within ${15 \mevcc}$ of the nominal \L mass~\cite{PDG2012}. To form \jpsi candidates, two oppositely-charged muons with $\pt(\mu)>0.5 \gevc$ are combined and their invariant mass is required to be within ${80 \mevcc}$ of the nominal $\jpsi$ mass.  Subsequently, \Lb candidates are formed by combining the \L and \jpsi candidates. To improve the \Lb mass resolution, the muons from the \jpsi decay are constrained to come from a common point and to have an invariant mass equal to the $\jpsi$ mass. We constrain the \L and \jpsi candidates to originate from a common vertex and to have an invariant mass between 5120 and ${6120 \mevcc}$. Moreover, \Lb candidates must have their momenta pointing to the associated PV by requiring ${\cos\theta_{\rm d}>0.99}$ where $\theta_{\rm d}$ is the angle between the \Lb momentum vector and the direction from the PV to the \Lb vertex. The associated PV is the PV having the smallest \ipchisq value.

To reduce the combinatorial background, a multivariate selection based on a boosted decision tree (BDT)~\cite{Breiman,AdaBoost} with eight variables is used. Five variables are related to the \Lb candidate: $\cos\theta_{\rm d}$, the \ipchisq, the proper decay time, the vertex $\chisq$ and the vertex separation \chisq between the PV and the vertex. Here, the vertex separation \chisq is the difference in \chisq between the nominal vertex fit and a vertex fit where the \Lb is assumed to have zero lifetime. The proper decay time is the distance between the associated PV and the \Lb decay vertex divided by the \Lb momentum. Two variables are related to the \jpsi candidate: the vertex $\chisq$ and the invariant mass of the two muons. The last variable used in the BDT is the invariant mass of the \L candidate. The BDT is using simulation for signal and sideband data ($M(\jpsi\L) > 5800 \mevcc$) for background in its training. The optimal BDT requirement is found separately for downstream and long candidates by maximising the signal significance $N_{\rm sig}/\sqrt{N_{\rm sig}+N_{\rm bkg}}$, where $N_{\rm sig}$ and $N_{\rm bkg}$ are the expected signal and background yields in a tight signal region around the \Lb mass. These two yields are estimated using the signal and background yields measured in data after the first set of loose requirements and using the BDT efficiency measured with the training samples. The BDT selection keeps about $90\%$ of the signal while removing about $80\%$ ($90\%$) of the background events for the downstream (long) candidates. Less background is rejected in the downstream case due to larger contamination from misreconstructed \BdToJPsiKS background decays. Candidates with $5550 < M(\jpsi\L) < 5700 \mevcc$ are used for the final analysis. In this mass range, the \BdToJPsiKS background is found to have a similar shape as the combinatorial background.

\section{Fitting procedure}
\label{sec:fitproc}

\begin{figure}[b!]
  \begin{center}
    \begin{tabular}{cc}
      \includegraphics[width=0.4\linewidth]{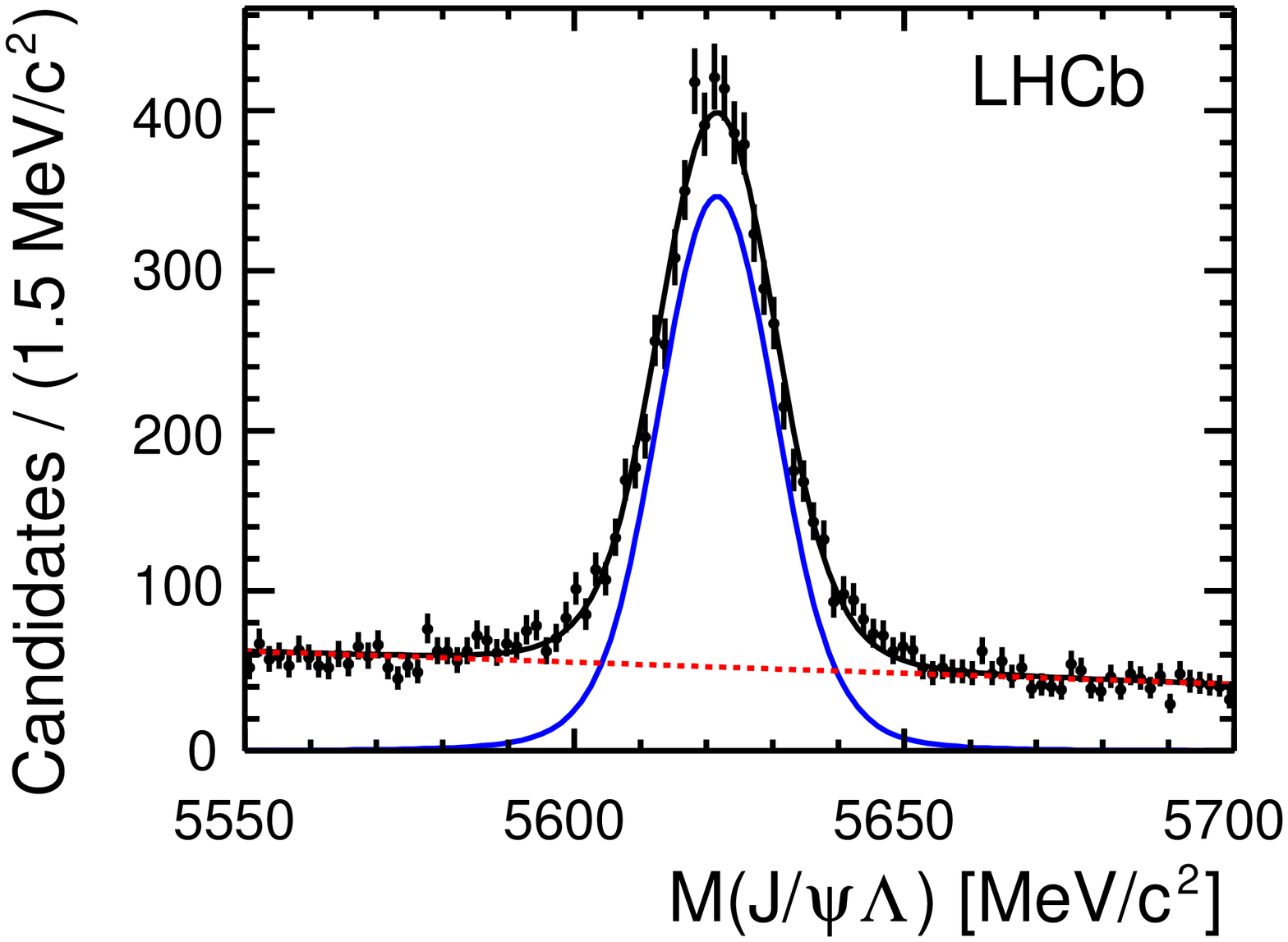} & 
      \includegraphics[width=0.4\linewidth]{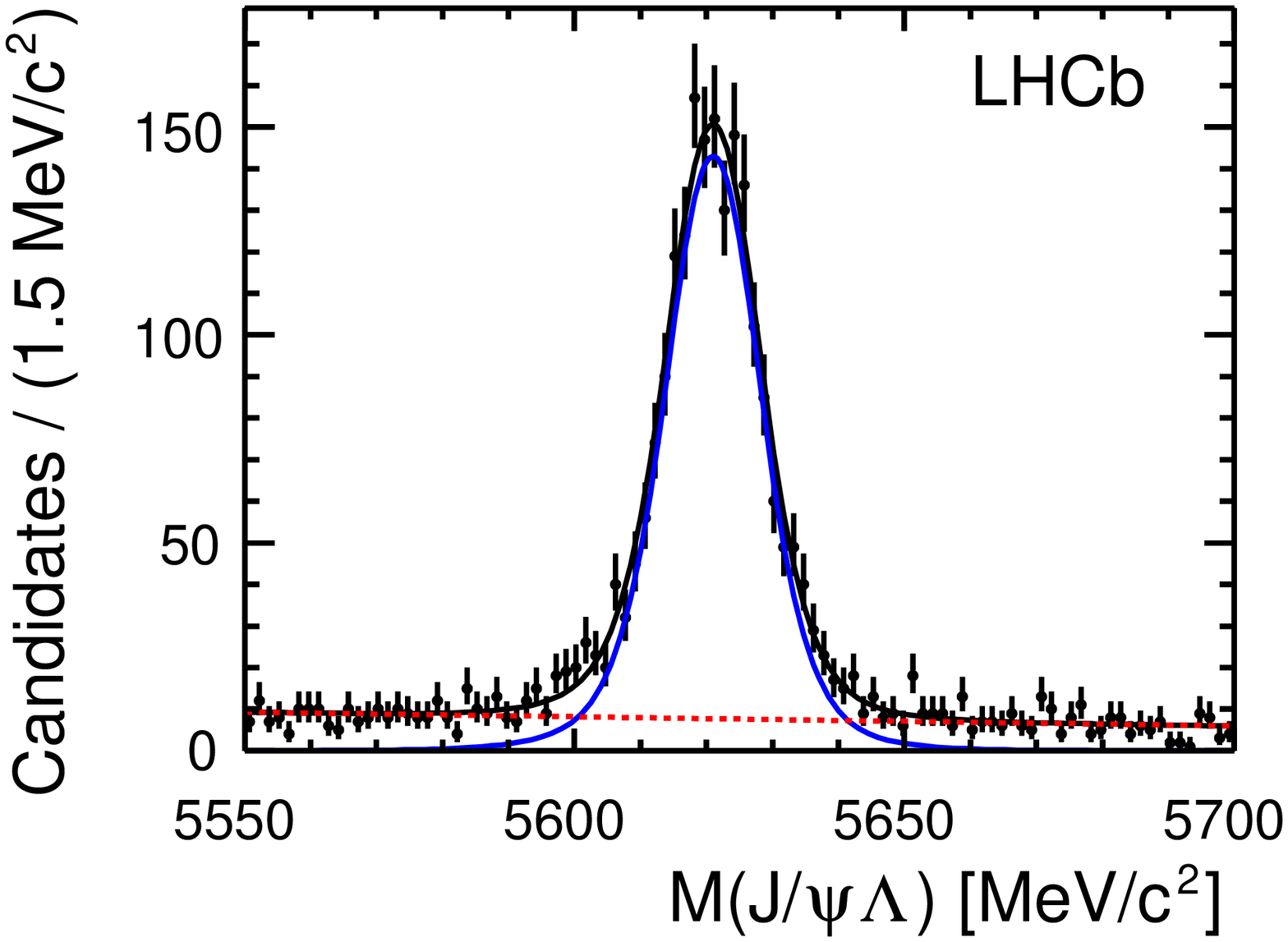} \\
    \end{tabular}
    \vspace*{-0.5cm}
  \end{center}
  \caption{
    \small
    Mass distribution for the \LbToLJPsi mode for the (left) downstream and (right) long candidates. The fitted signal component is shown as a solid blue curve while the background component is shown as a dashed red line.}
  \label{fig:lbmass}
\end{figure}

An unbinned extended maximum likelihood fit to the mass distribution of the \Lb candidates is performed. The likelihood function is defined as
\begin{equation}
\mathcal{L}_{\rm mass} = \frac{e^{-\sum_j N_j}}{N!} \times \prod_{i=1}^{N} \left ( \sum_j N_j P_j(M_i(\jpsi\L)) \right),
\end{equation}
where $i$ runs over the events, $j$ runs over the different signal and background probability density functions (\PDF), $N_j$ are the yields and $P_j$ the \PDFs. The sum of two Crystal Ball functions~\cite{Skwarnicki:1986xj} with opposite side tails and common mean and width parameters is used to describe the signal mass distribution. The mean and width parameters are left free in the fit while the other parameters are taken from the simulated signal sample. The background is modelled with a first-order polynomial function. The candidates reconstructed from downstream and long \L combinations are fitted separately taking into account that the resolution is worse for the downstream signal candidates. The results of the fits to the mass distributions are shown in Fig.~\ref{fig:lbmass}. We obtain $5346 \pm 96$ ($5189 \pm 95$) downstream and $1861 \pm 49$ ($761 \pm 36$) long signal (background) candidates. Using the results of this fit, {\sWeight}s (\wmass) are computed by means of the \sPlot technique~\cite{splot}, in order to statistically subtract the background in the angular distribution.

To ensure accurate modelling of the signal, corrections to the \pt and rapidity~($y$) spectra are obtained by comparing the simulation with data by means of the \sPlot technique. For the \Lb and \L particles, the simulated data is corrected using two-dimensional $(\pt,y)$ distributions in order to better reproduce the data. These distributions do not depend on the polarisation and the decay amplitudes but have an impact on the reconstruction acceptance. The same procedure is used on the pion of \BdToJPsiKS decays and is subsequently used to calibrate the $(\pt,y)$ spectrum of the pion of the \LbToLJPsi decay.

\begin{figure}[tb!]
  \begin{center}
    \begin{tabular}{ccc}
      \includegraphics[width=0.3\linewidth]{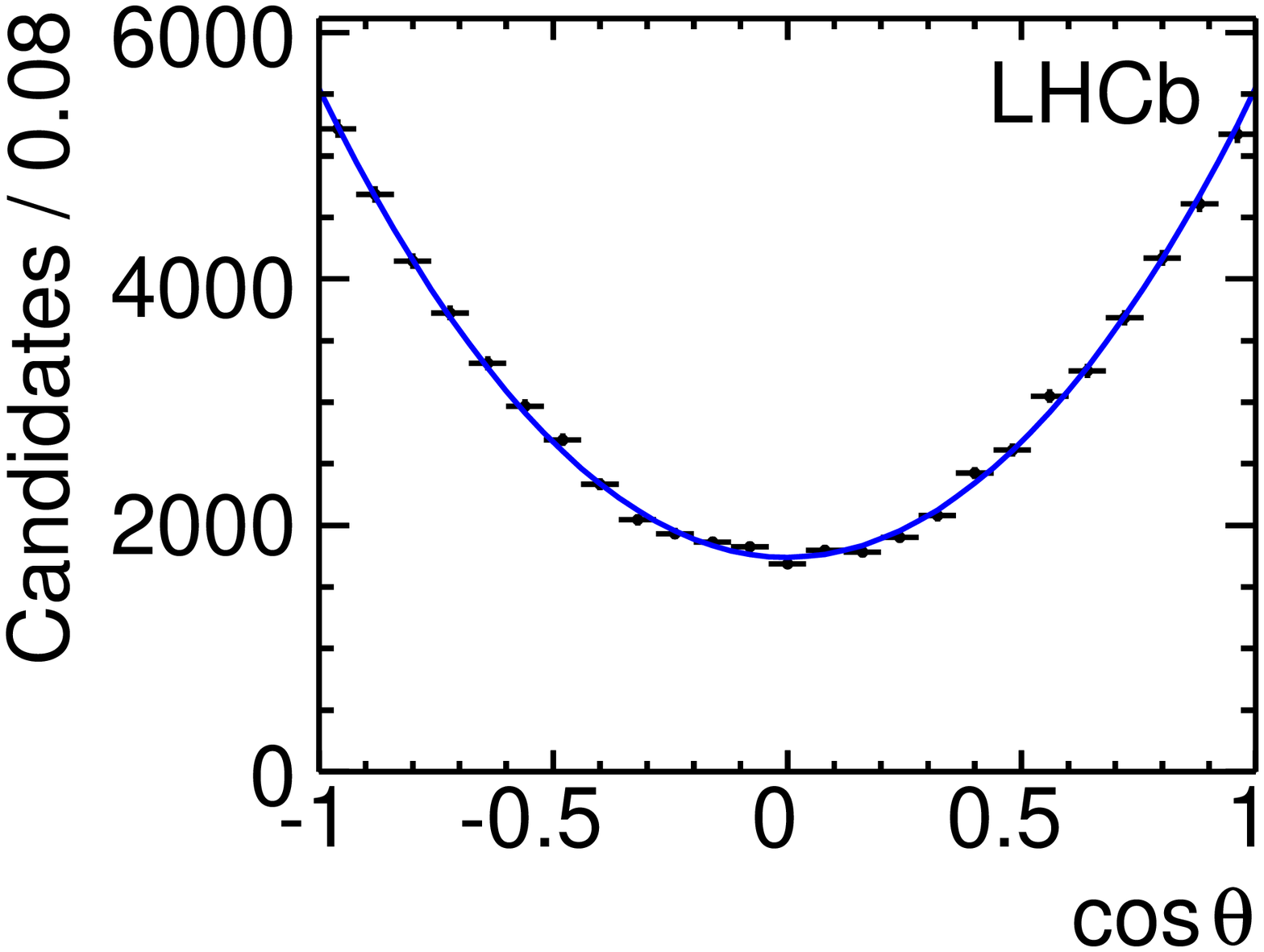} &
      \includegraphics[width=0.3\linewidth]{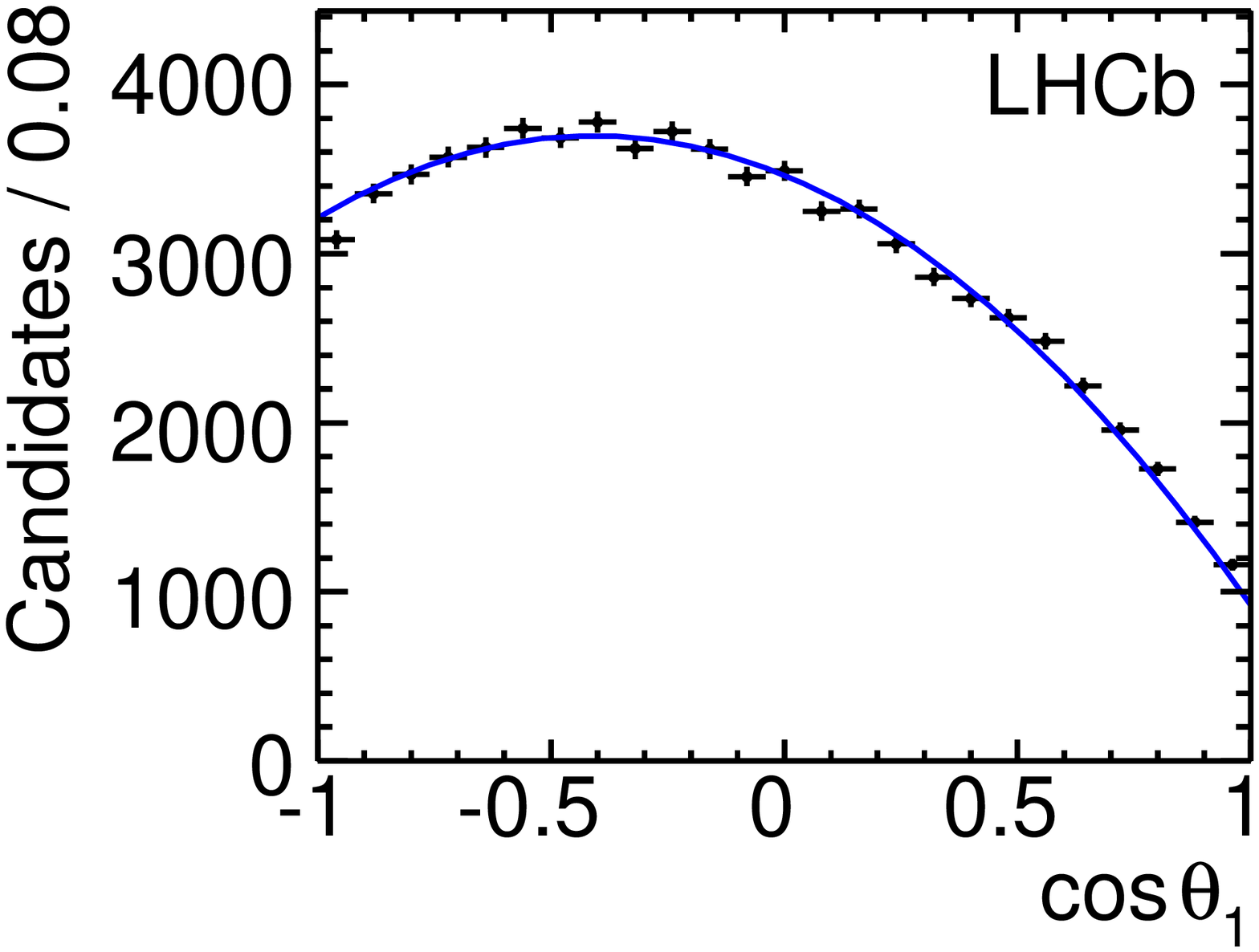} &
      \includegraphics[width=0.3\linewidth]{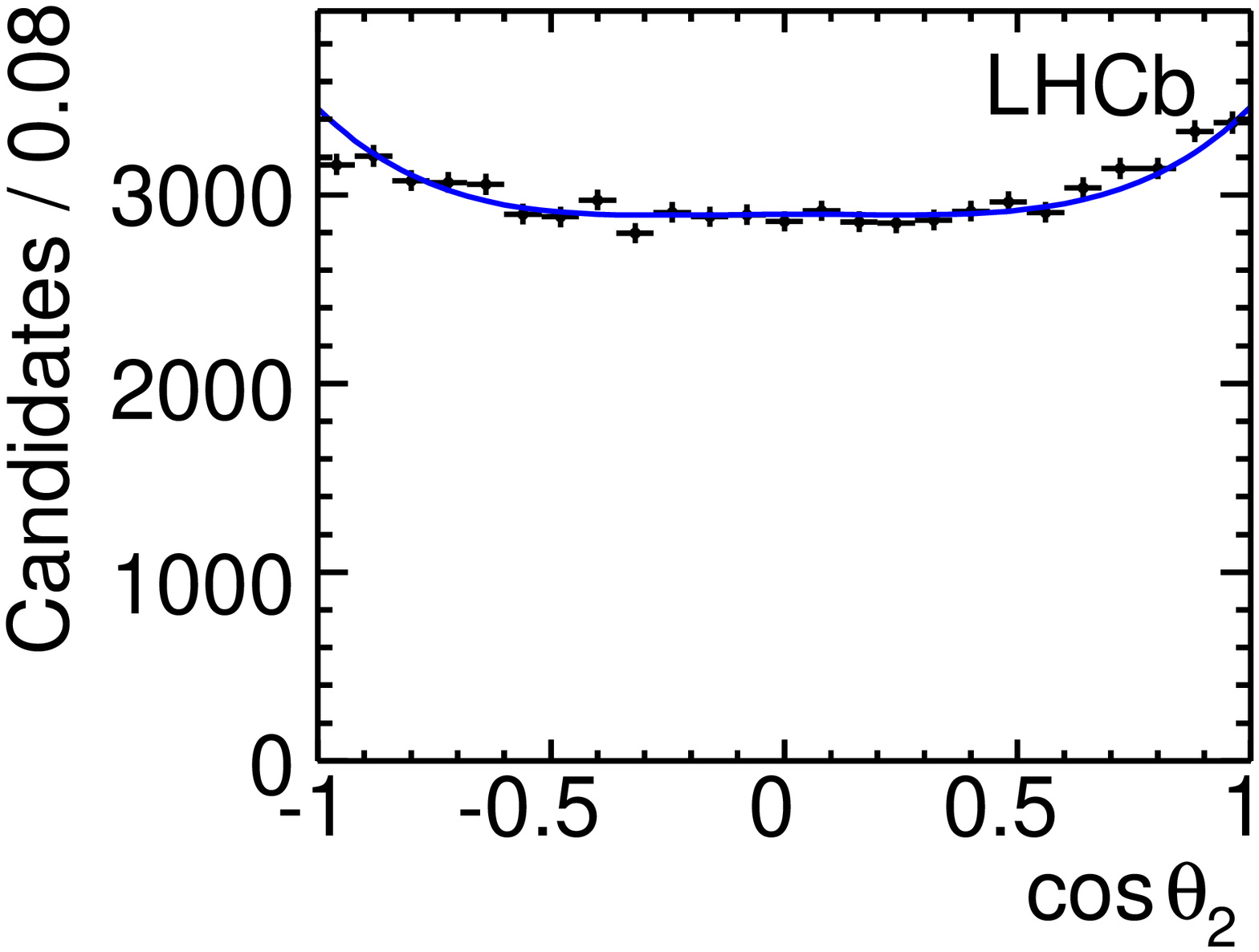} \\
      \includegraphics[width=0.3\linewidth]{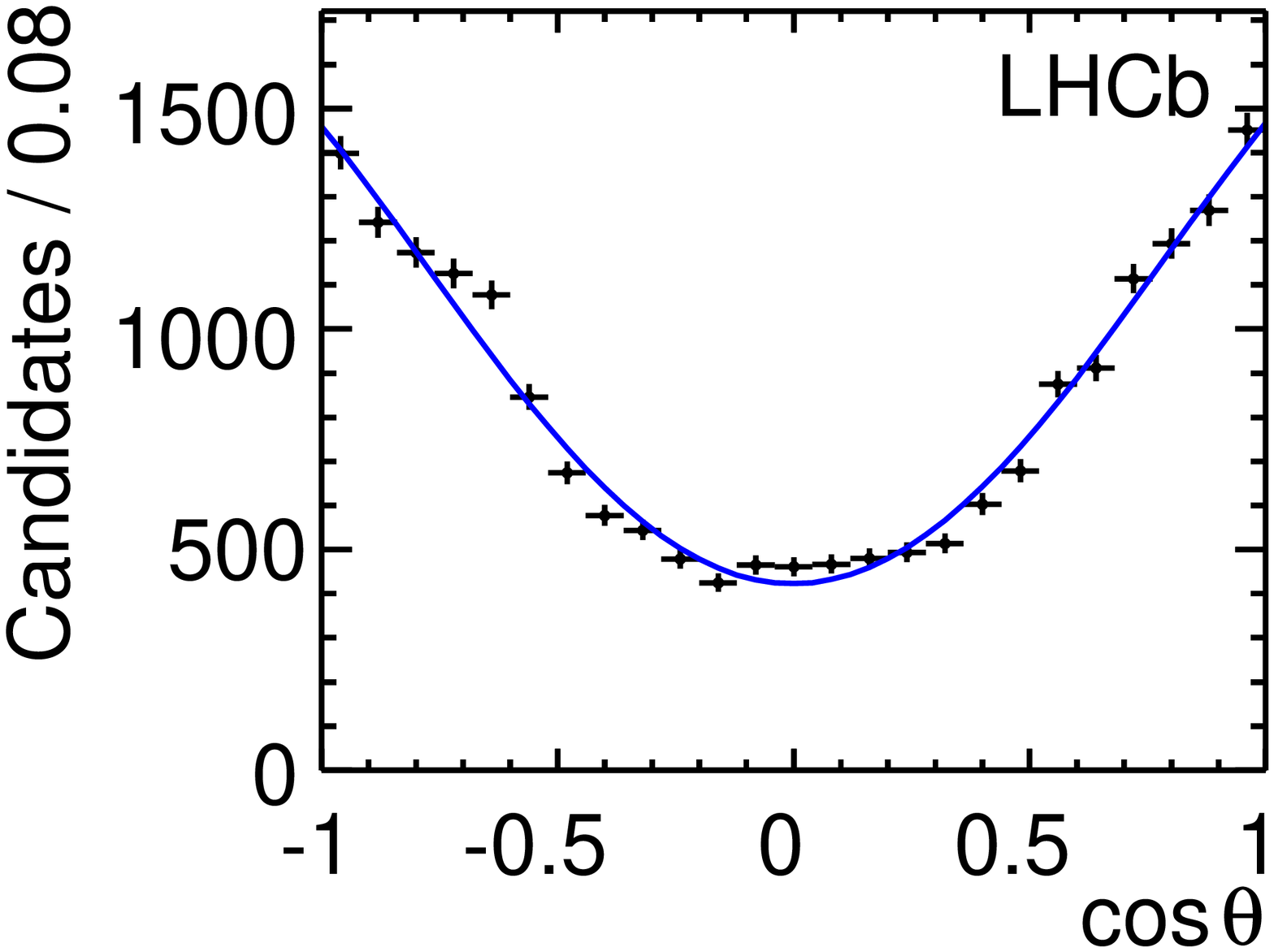} &
      \includegraphics[width=0.3\linewidth]{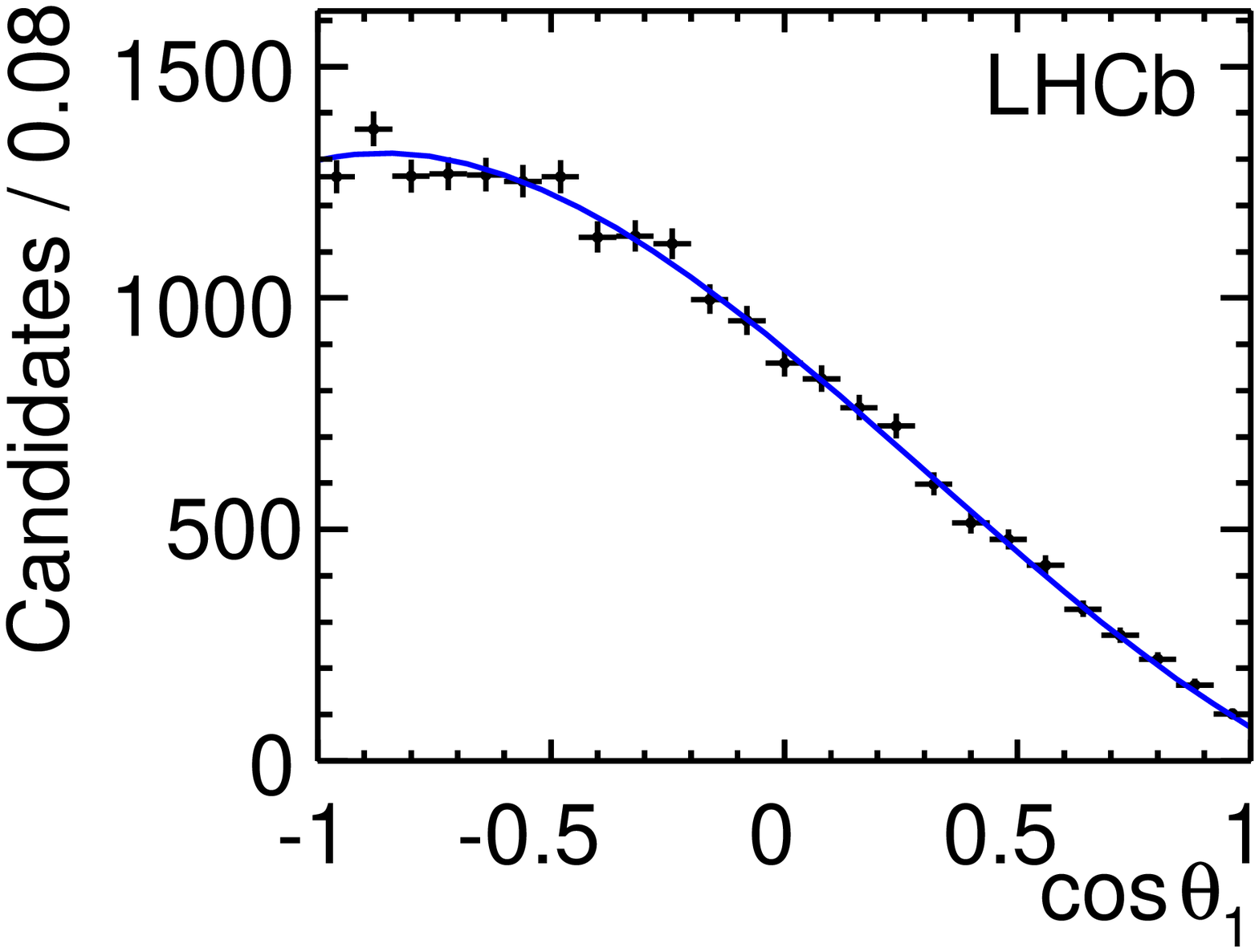} &
      \includegraphics[width=0.3\linewidth]{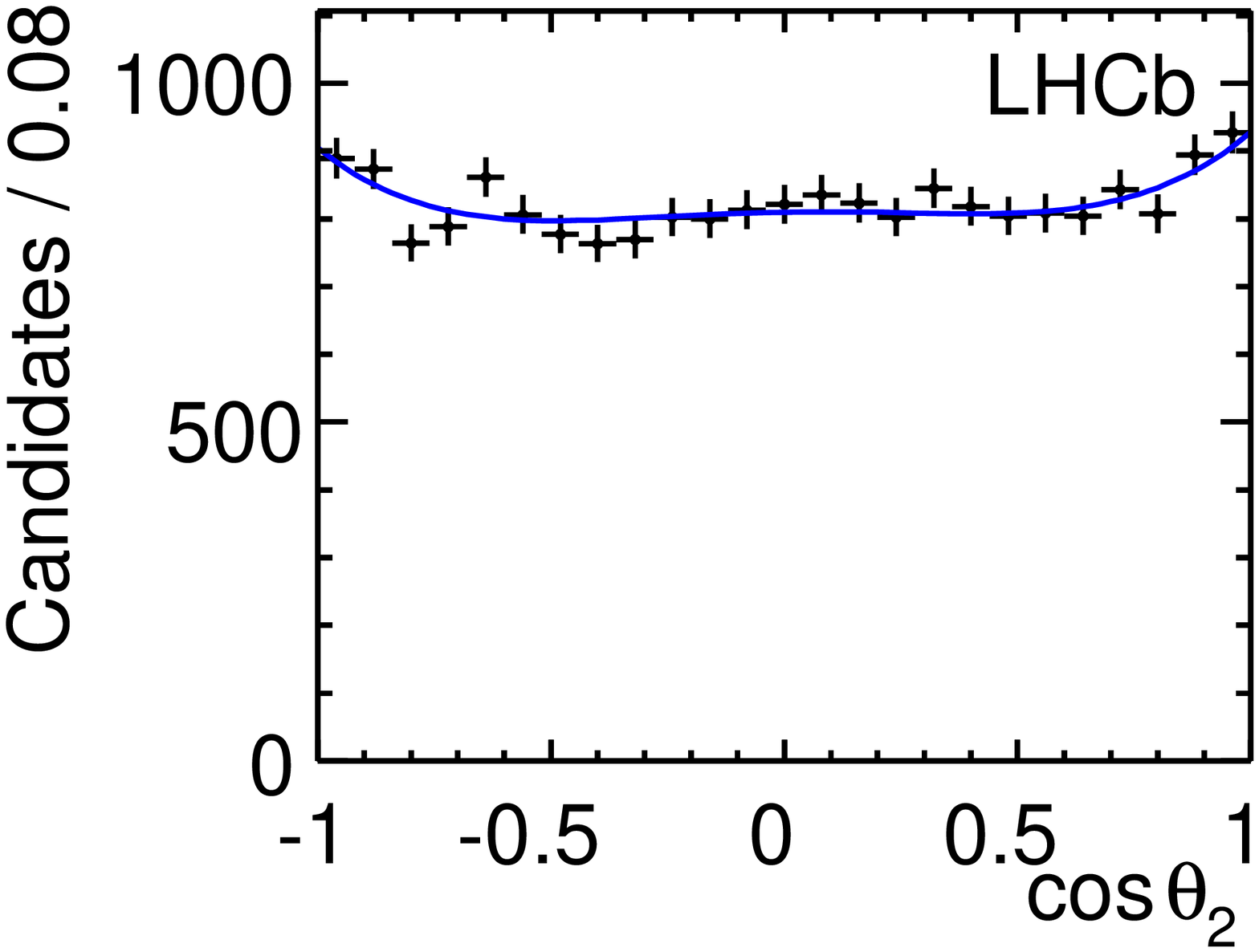} \\
    \end{tabular}
    \vspace*{-0.5cm}
  \end{center}
  \caption{
    \small
    Projections of the acceptance function together with the simulated signal data for (top) downstream and (bottom) long candidates.}
  \label{fig:acclb}
\end{figure}

Since the detector acceptance depends on the three decay angles, the acceptance is modelled with a sum of products of Legendre polynomials ($L_i$)
\begin{equation}
f_{\rm acc} = \sum_{i,j,k} c_{ijk} L_i(\cthz) L_j(\ctho) L_k(\ctht),
\label{eqn:acc}
\end{equation}
where $i$ and $k$ are chosen to be even or equal to one. Unbinned maximum likelihood fits to the simulated signal candidates are performed, separately for downstream and long candidates. The simulated is produced using a phase-space model and unpolarised \Lb baryons. The three angular distributions are therefore uniformly generated. Acceptances of the \Lb and \Lbbar decays are found to be statistically consistent. A common acceptance function is therefore used. The maximum orders of the Legendre polynomials are chosen by comparing the fit probability. The requirements $i<5$, $j<4$, $k<5$ and $i+j+k<9$ are chosen. The results of the fit to the acceptance distributions are shown in Fig.~\ref{fig:acclb}. 

We then perform an unbinned likelihood fit to the (\cthz, \ctho, \ctht) distribution. Each candidate is weighted with ${\wtot = \wmass \times \wacc}$ where \wmass subtracts the background and ${\wacc = 1/f_{\rm acc}(\cthz, \ctho, \ctht)}$ corrects for the angular acceptance~\cite{sfit}. The sum of the \wmass weights over all the events is by construction equal to the signal yield, and \wtot is normalised in the same way. Since the weighting procedure performs background subtraction and corrects for acceptance effects, only the signal \PDF has to be included in the fit of the angular distribution. The detector resolution is neglected in the nominal fit as it is found to have little effect on the results. It will be considered as source of systematic uncertainty. The likelihood is therefore
\begin{equation}
\mathcal{L}_{\rm ang} =  \prod_{i=1}^{N} \wtot^{i} \dGdomegat(\cthz^i, \ctho^i, \ctht^i),
\label{eqn:lang}
\end{equation}
where $i$ runs over all events. A simultaneous fit to the angular distributions of the downstream and long samples is performed. The \al parameter is fixed to its measured value, $0.642 \pm 0.013$~\cite{PDG2012}.

The accurate modelling of the acceptance is checked with a similar decay, \BdToJPsiKS. Here, the angular distribution is known, and \Bz mesons are unpolarised. These decays are selected in the same way as signal, and the fitting procedure described above is performed. Agreement with the expected (\cthz, \ctho, \ctht) distribution is obtained.

\section{Results}
\label{sec:results}

\begin{figure}[tb!]
  \begin{center}    
    \begin{tabular}{ccc}
      \includegraphics[width=0.31\linewidth]{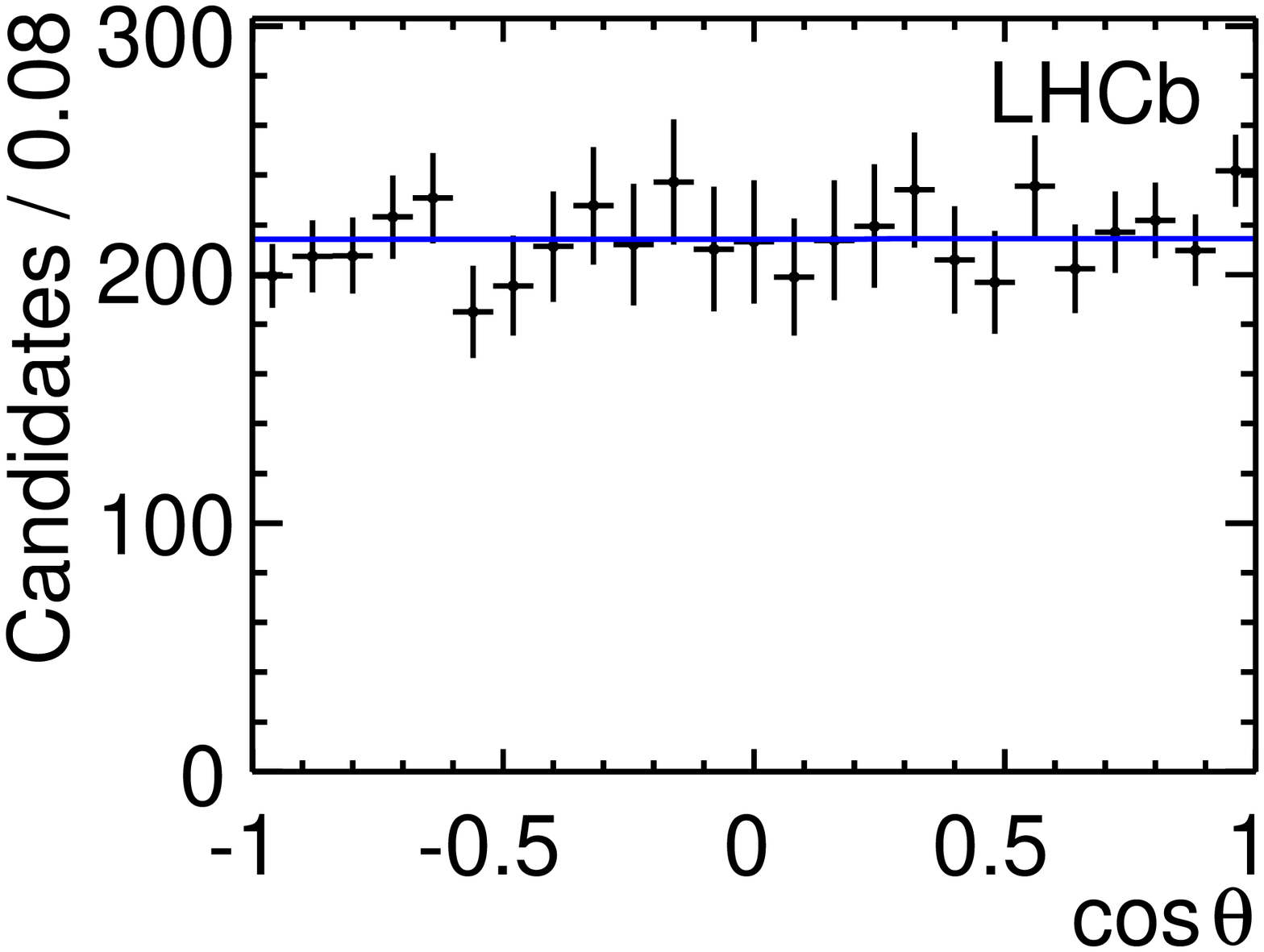} &
      \includegraphics[width=0.31\linewidth]{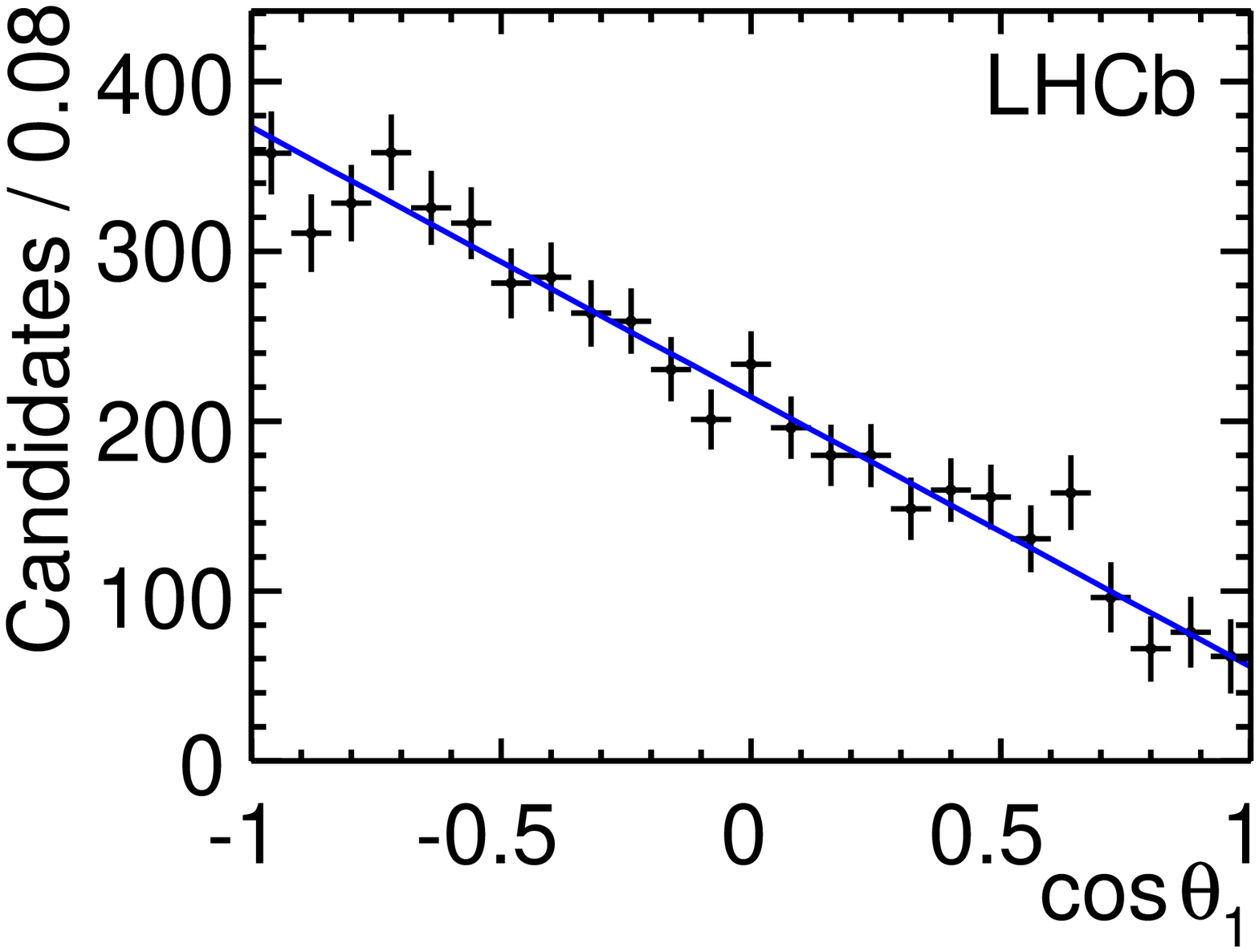} &
      \includegraphics[width=0.31\linewidth]{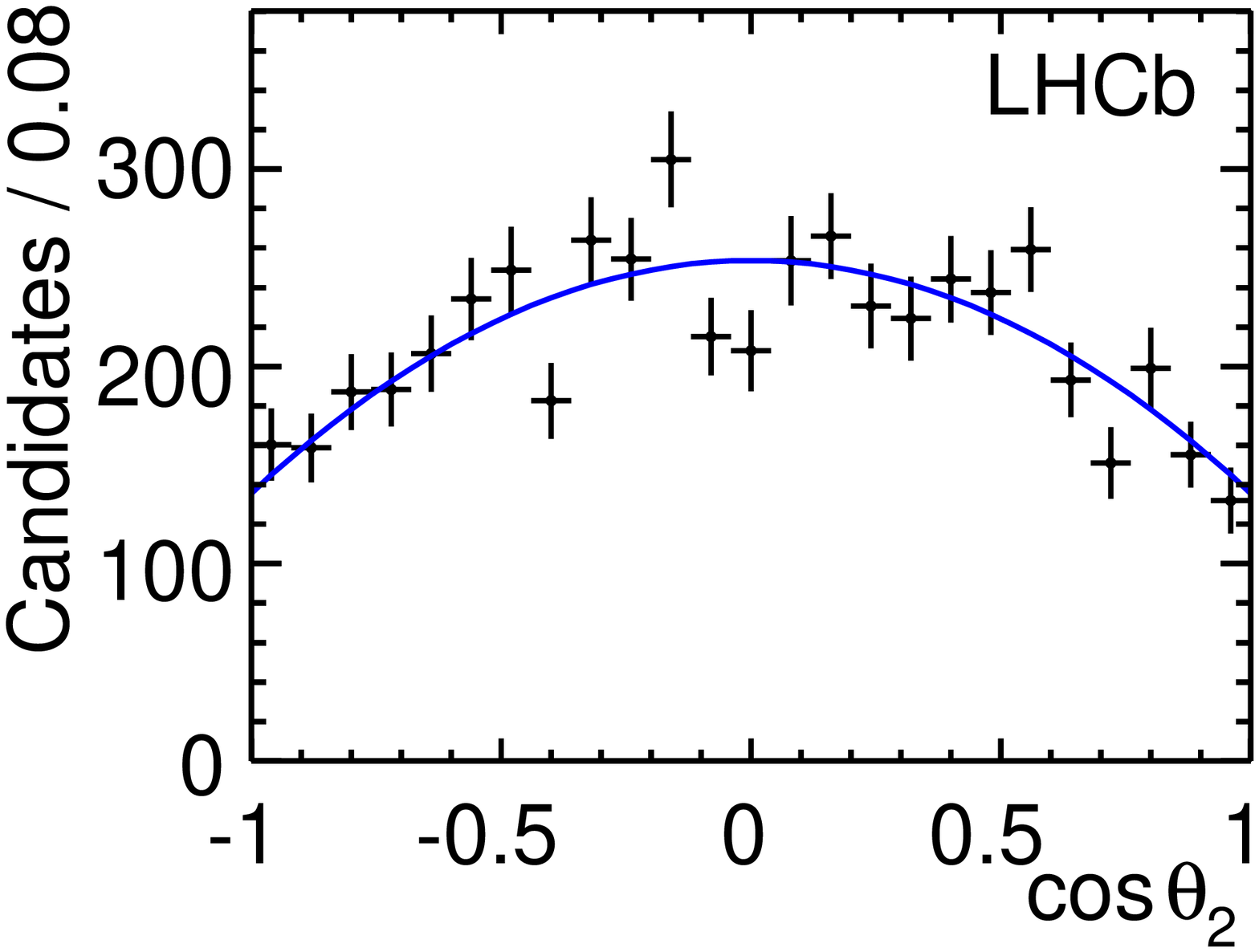} \\
      \includegraphics[width=0.31\linewidth]{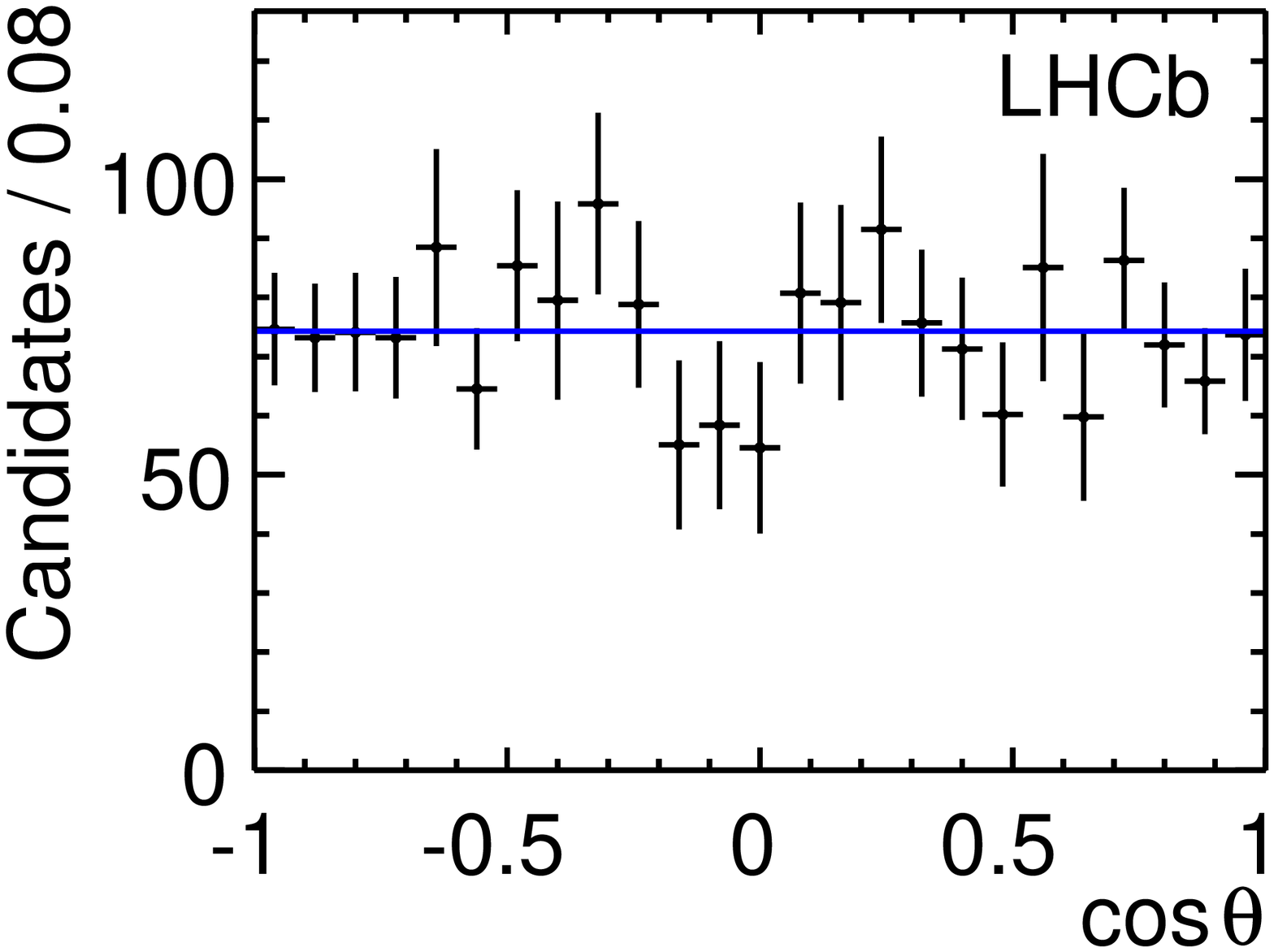} &
      \includegraphics[width=0.31\linewidth]{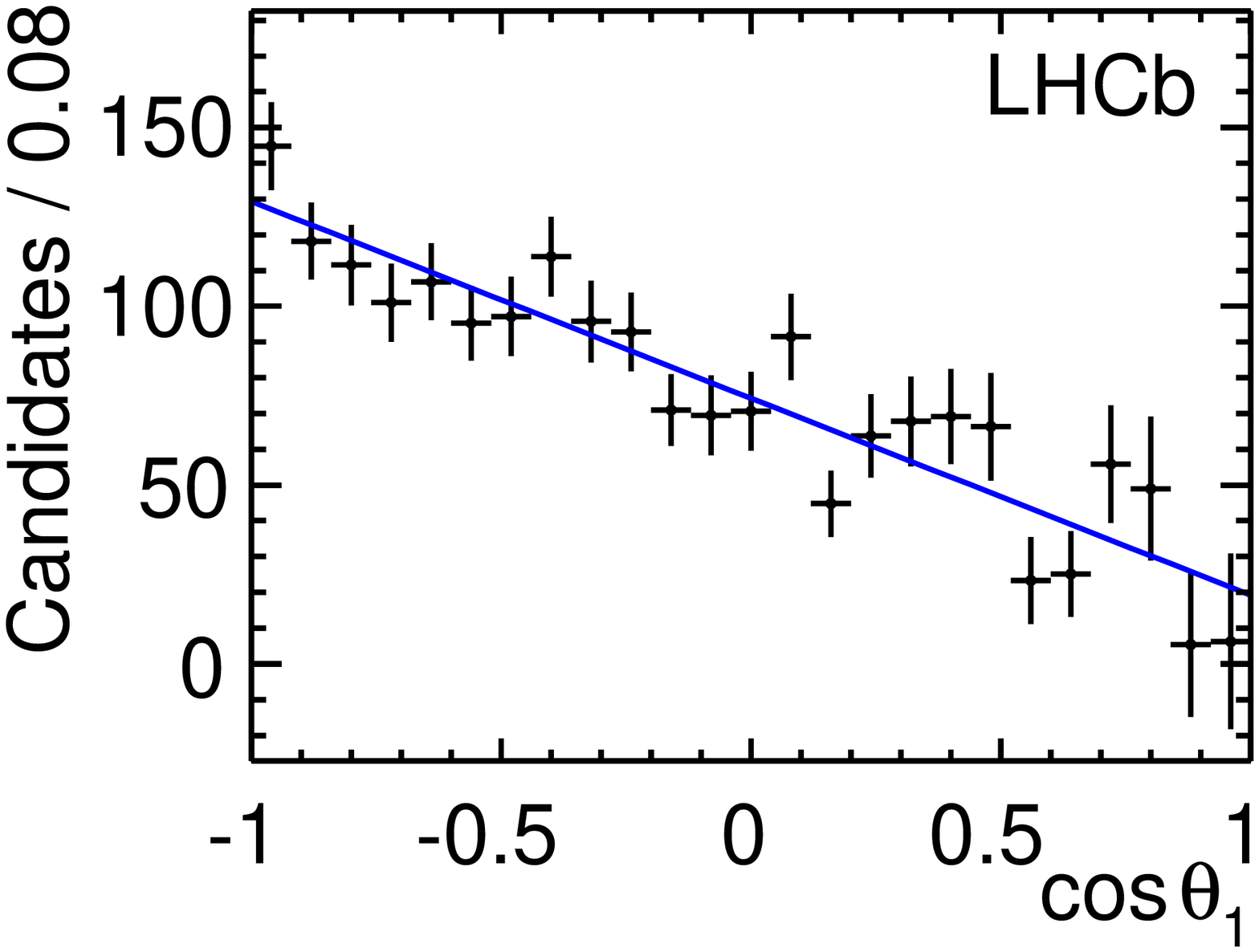} &
      \includegraphics[width=0.31\linewidth]{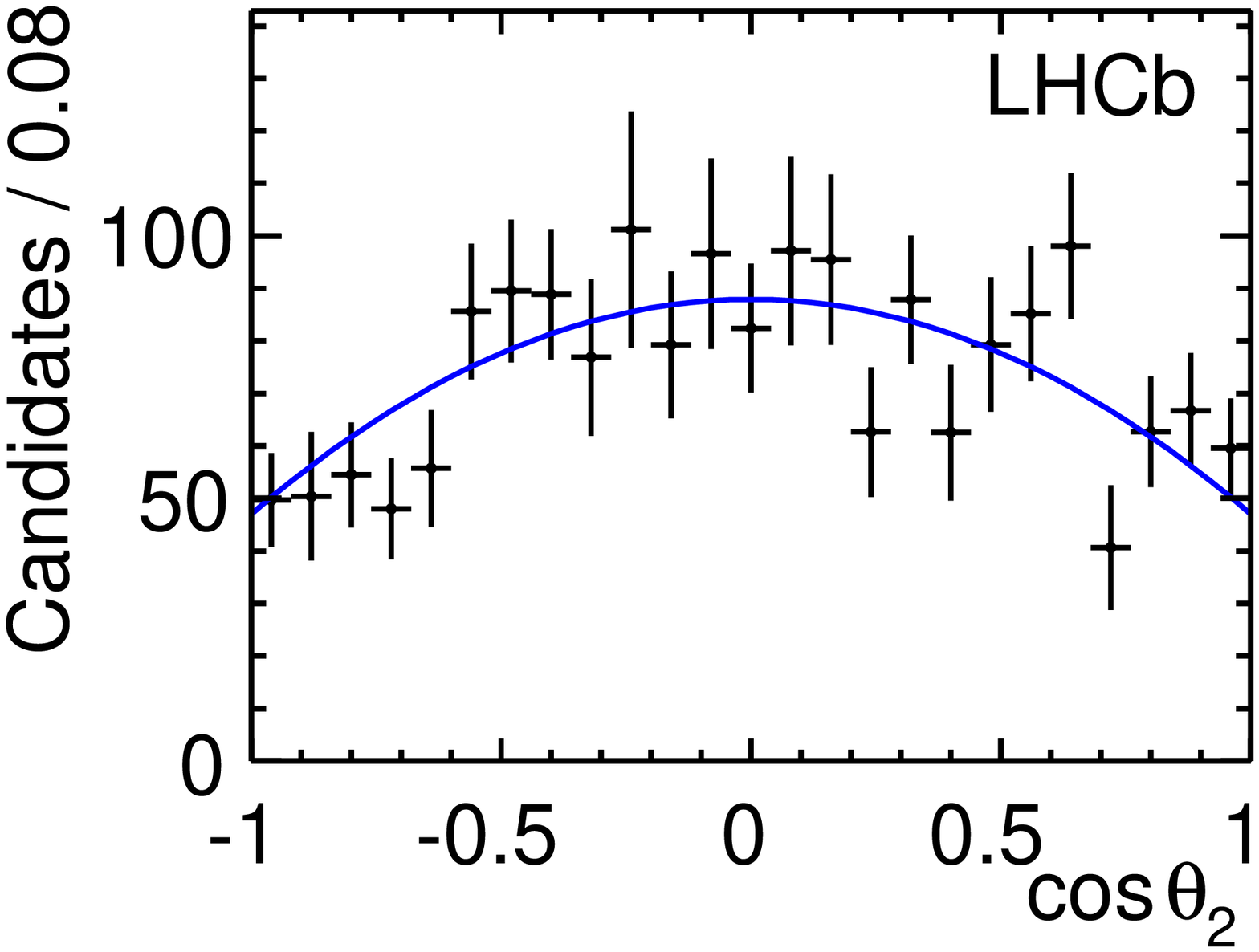} \\
    \end{tabular}
    \vspace*{-0.5cm}
  \end{center}
  \caption{
    \small
    Projections of the angular distribution of the background subtracted and acceptance corrected \LbToLJPsi data for the (top) downstream and (bottom) long candidates. The fit is shown as solid lines.}
  \label{fig:lbang}
\end{figure}

The results of the fits to the angular distributions of the weighted \LbToLJPsi data are shown in Fig.~\ref{fig:lbang}. We obtain the following results: $\Polb = 0.06 \pm 0.06$, $\ab = 0.00 \pm 0.10$, $\rz = 0.58 \pm 0.02$ and $\ro = -0.58 \pm 0.06$, where the uncertainties are statistical only.

The polarisation could be different between \Lb and \Lbbar due to their respective production mechanisms. The data are separated according to the \Lb flavour and fitted using the same amplitude parameters but different parameters for the \Lb and \Lbbar polarisations. As compatible results are obtained within statistical uncertainties, the polarisations of \Lb and \Lbbar baryons are assumed to be equal.

A possible bias is investigated by fitting samples of generated  experiments with sizes and parameters close to those measured in data. We generate many samples varying \ab between $-0.25$ to $0.25$ while keeping \rz equal to $-\ro$, thus keeping $\mbmt$ and $\mapt$ equal to zero. We find that the fitting procedure biases all parameters toward negative values, slightly for $\Polb$ and $\rz$ ($\sim$$10\%$ of their respective statistical uncertainties) and more significantly for $\ab$ and $\ro$ ($\sim$$40\%$ of their respective statistical uncertainties). For \Polb and \rz, the biases do not change significantly when changing the value of \ab used to generate the simulated samples. On the other hand, the biases on \ab and \ro do change, and the observed discrepancies are treated as systematic uncertainties. Moreover, the statistical uncertainties on the four fit parameters are underestimated: again slightly for $\Polb$ and $\rz$ and significantly, by a factor of $\sim$$1.7$, for $\ab$ and $\ro$. 

We correct the measured values and statistical uncertainties of the four fit parameters. The corrected statistical uncertainties are obtained by multiplying the covariance matrix with a correction matrix obtained from the study of the simulated samples. This correction matrix contains on its diagonal the squares of the widths of the pull distributions of the four fit parameters. The remaining entries of this matrix are set to zero as the correlation matrix computed with the results of the fits of the generated samples is found to be very close to the correlation matrix calculated when fitting the data.

Finally, the corrected result is $\Polb =  0.06 \pm 0.07$, $\ab = 0.05 \pm 0.17$, $\rz = 0.58 \pm 0.02$, $\ro = -0.56 \pm 0.10$, where the uncertainties are statistical only. The corrected statistical correlation matrix between the four fit parameters (\Polb, \ab, \rz, \ro) is
\begin{align*}
\begin{pmatrix*}[l]
 1 & \phantom{-}0.10  & -0.07        & \phantom{-}0.13 \\ 
   & \phantom{-}1     & -0.63        & \phantom{-}0.95 \\
   &                  & \phantom{-}1 & -0.56 \\
   &                  &              & \phantom{-}1 
\end{pmatrix*}.
\end{align*}
Large correlations are not seen between the polarisation and the amplitude parameters. On the other hand, the amplitude parameters are strongly correlated with respect to each other, \ab and \ro being almost fully correlated.

\section{Systematic uncertainties and significance}
\label{sec:syst}

The systematic uncertainty on each measured physics parameter is evaluated by repeating the fit to the data varying its input parameters assuming Gaussian distributions and taking into account correlations when possible. The systematic uncertainties are summarised in Table~\ref{tab:systematics}. They are dominated by the uncertainty arising from the acceptance function, the calibration of the simulated signal sample and the fit bias. The uncertainty related to the acceptance function is obtained by varying the coefficients of the Legendre function within their uncertainties and taking into account their correlations. For the calibration of our simulated data, the uncertainty is obtained when changing the $(\pt,y)$ calibrations of the \Lb, \L and pion particles within their uncertainties and obtaining a new acceptance function. The function that is used to fit the data does not include the effect of the angular resolution. The angular resolution, obtained with simulated samples, is negligible for \thz and \tht. However, it is large, up to $\sim$$70\%$, for small values of \tho. The systematic uncertainty is obtained by fitting simulated samples in which the resolution effect is introduced. Effects of the deviation from an uniform acceptance in \phio and \phit assumed in Eq.~(\ref{eqn:w3}) are found to be negligible. The simplification to use only one component to describe the background is found not to bias the result. Other systematic uncertainties are small or negligible. These are related to the signal mass PDF parameters, the background subtraction and \al. The uncertainty related to the background subtraction are obtained when varying the obtained result of the mass fit and computing the \wmass weights again. The \al parameter is varied within its measurement uncertainties~\cite{PDG2012}.

\begin{table}[tb!]
  \caption{Absolute systematic uncertainties on the measured parameters.}
  \begin{center}
    \newcolumntype{.}{D{.}{.}{-1}}
    \begin{tabular}{lllll}
      Source                      & \Polb & \phantom{$<$}\ab    & \phantom{$<$}\rz    & \ro   \\ \hline 
      Acceptance                  & 0.02  & \phantom{$<$}0.04   & \phantom{$<$}0.006  & 0.03  \\ 
      Simulated data calibration  & 0.01  & \phantom{$<$}0.04   & \phantom{$<$}0.006  & 0.03 \\ 
      Fit bias               & 0.004 & \phantom{$<$}0.04   & \phantom{$<$}0.001  & 0.02 \\ 
      Angular resolution     & 0.002 & \phantom{$<$}0.01 & $<$0.001 & 0.005 \\ 
      Background subtraction & 0.001 & \phantom{$<$}0.006  & \phantom{$<$}0.001  & 0.005 \\ 
      \al                    & 0.002 & $<$0.001  & $<$0.001  & 0.01  \\ \hline
      Total (quadratic sum)         & 0.02  & \phantom{$<$}0.07    & \phantom{$<$}0.01  & 0.05  \\
    \end{tabular}
  \end{center}
  \label{tab:systematics}
\end{table}

To compare our results with a prediction on a parameter $p$, we compute the significance with respect to a $p_{\rm test}$ value using a profile along $p$ of the likelihood function, \ie\ the likelihood value obtained when varying $p$ and minimising with respect to the other parameters. A Monte Carlo integration is performed to include the systematic uncertainties in the likelihood profiles. We perform the fit to the data when varying all systematic uncertainties and obtain a likelihood profile for each fit of the data. The likelihood profile which includes all systematic uncertainties is then the average of all the obtained profiles. The significance is defined as $\mathcal{S}(p=p_{\rm test}) = \sqrt{2 (\log \mathcal{L}(p_{\rm test}) - \log \mathcal{L}(p_0))}$, where $\mathcal{L}(p_0)$ is the likelihood value of the nominal fit. Significances are given in the concluding section of this Letter.

\section{Conclusion}
\label{sec:conclusion}

We have performed an angular analysis of about 7200 \LbToLppiJPsimumu decays. The \LbToLJPsi decay amplitudes are measured for the first time, and the \Lb production polarisation for the first time at a hadron collider. The results are
\begin{align*}
\Polb &= \phantom{-}0.06 \pm 0.07 \pm 0.02, \\
\ab   &= \phantom{-}0.05 \pm 0.17 \pm 0.07, \\
\rz   &=  \phantom{-}0.58 \pm 0.02 \pm 0.01, \\
\ro   &= -0.56 \pm 0.10 \pm 0.05,
\end{align*}
which correspond to the four helicity amplitudes
\begin{align*}
\mapt &= \phantom{-} 0.01 \pm 0.04 \pm 0.03, \\
\mamt &= \phantom{-} 0.57 \pm 0.06 \pm 0.03, \\
\mbpt &= \phantom{-} 0.51 \pm 0.05 \pm 0.02, \\
\mbmt &= -0.10 \pm 0.04 \pm 0.03,
\end{align*}
where the first uncertainty is statistical and the second systematic. The reported polarisation and amplitudes are obtained for the combination of \Lb and \Lbbar decays. More data are required to probe any possible difference.

Our result cannot exclude a transverse polarisation at the order of $10\%$~\cite{Hiller:2007ur}. However, a value of $20\%$ as mentioned in Ref.~\cite{Ajaltouni:2004zu} is disfavoured at the level of $2.7$ standard deviations.

For the \Lb asymmetry parameter, our result is compatible with the predictions ranging from $-21\%$ to $-10\%$~\cite{Cheng:1996cs,Fayyazuddin:1998ap,Mohanta:1998iu,Chou:2001bn,Wei:2009np} but does not agree with the HQET prediction of $77.7\%$~\cite{Ajaltouni:2004zu} at $5.8$ standard deviations.

\section*{Acknowledgements}

\noindent We express our gratitude to our colleagues in the CERN
accelerator departments for the excellent performance of the LHC. We
thank the technical and administrative staff at the LHCb
institutes. We acknowledge support from CERN and from the national
agencies: CAPES, CNPq, FAPERJ and FINEP (Brazil); NSFC (China);
CNRS/IN2P3 and Region Auvergne (France); BMBF, DFG, HGF and MPG
(Germany); SFI (Ireland); INFN (Italy); FOM and NWO (The Netherlands);
SCSR (Poland); ANCS/IFA (Romania); MinES, Rosatom, RFBR and NRC
``Kurchatov Institute'' (Russia); MinECo, XuntaGal and GENCAT (Spain);
SNSF and SER (Switzerland); NAS Ukraine (Ukraine); STFC (United
Kingdom); NSF (USA). We also acknowledge the support received from the
ERC under FP7. The Tier1 computing centres are supported by IN2P3
(France), KIT and BMBF (Germany), INFN (Italy), NWO and SURF (The
Netherlands), PIC (Spain), GridPP (United Kingdom). We are thankful
for the computing resources put at our disposal by Yandex LLC
(Russia), as well as to the communities behind the multiple open
source software packages that we depend on.

\addcontentsline{toc}{section}{References}

\bibliographystyle{LHCb}
\bibliography{main}

\end{document}